\documentclass[12 pt]{article}
\usepackage[utf8]{inputenc}
\usepackage{authblk}
\usepackage{epigraph}
\usepackage{hyperref}
\usepackage{graphicx}
\usepackage{mathtools}   % loads »amsmath«
\usepackage{amsfonts}
\usepackage{appendix}
\usepackage{caption}
\usepackage{orcidlink}
\graphicspath{ {./figures/} }
\hypersetup{
    colorlinks=true,
    linkcolor=blue,
    urlcolor=blue,
}
\providecommand{\keywords}[1]
{
  \small	
  \textbf{\textit{Keywords---}} #1
}

\title{The problem of time for non-deparametrizable models and quantum gravity\thanks{Cite as: Mozota Frauca, \'A. The Problem of Time for Non-Deparametrizable Models and Quantum Gravity (2024), In F. Bianchini, V. Fano and P. Graziani (Eds.), Current Topics in Logic and the Philosophy of Science (College Publications).}}

\author[1,2]{Álvaro Mozota Frauca}
\affil[ ]{alvaro.mozota@upc.edu, \orcidlink{0000-0002-7715-0563} \href{https://orcid.org/0000-0002-7715-0563}{https://orcid.org/
0000-0002-7715-0563}}
\affil[1]{Department of Mathematics. Universitat Polit\`ecnica de Catalunya. Barcelona (Spain)}
\affil[2]{Department of Condensed Matter Physics.Universitat de Barcelona. Barcelona (Spain)}

\date{February 12, 2024}

\begin{document}

\maketitle

\begin{abstract}
In this article I introduce a distinction between two types of reparametrization invariant models and I argue that while both suffer from a problem of time at the time of applying canonical quantization methods to quantize them, its severity depends greatly on the type of model. Deparametrizable models are models that have time as a configuration or phase space variable and this makes it the case that the problem of time can be solved. In the case of non-deparametrizable models, we cannot find time in the configuration or phase space of the model, and hence the techniques that allow solving the problem in the deparametrizable case do not apply. This seems to signal that the canonical quantization techniques fail to give a satisfactory quantization of non-deparametrizable models. As I argue that general relativity is non-deparametrizable, this implies that the canonical quantization of this theory may fail to provide a successful theory of quantum gravity. \\
\end{abstract}

\keywords{quantum gravity, problem of time, general relativity, canonical quantization}

\section{Introduction}

One of the strategies that physicists have deployed in order to build a theory of quantum gravity is to apply canonical quantization techniques to general relativity. However, it is a well-known fact that the application of these techniques leads to what is known as the problem of time and, hence, that it affects all the approaches to quantum gravity that are based on this strategy\footnote{Other approaches, such as string theory, are based on completely different ideas and techniques and avoid this problem.}. In a nutshell, one follows the same quantization techniques that are used for quantizing gauge theories and what one obtains is that the supposedly physical states lack any temporal dependence. This has been known from the first quantization of general relativity by Wheeler and deWitt \cite{DeWitt1967b} and discussed since. Even though the two most comprehensive discussions and reviews of the problem in the 90s \cite{Kuchar1992,Isham1993} show some of the serious difficulties associated with this problem and the shortcomings of some of the resolutions proposed, the quantum gravity community has since adopted the view that the problem of time can be overcome in one way or another and that canonical quantization methods, when applied to general relativity, produce a meaningful theory, even if apparently timeless\footnote{See for instance the discussions in \cite{Anderson2017,Kiefer2012,Rovelli2004,Rovelli2015}.}.

Some authors have been critical of this view\footnote{See \cite{Barbour1994,Gryb2010,Gryb2014,Gryb2016}.}, and in particular I have recently argued \cite{MozotaFrauca2023} that there are good reasons for believing that the canonical quantization of general relativity fails to give a satisfactory quantum theory. In this article I will expand on this criticism by leaving aside some of the more technical details and focusing on its more conceptual part.

I will start first in section \ref{sect_gauge} by introducing reparametrization invariance as a symmetry that physical models in general can have and that general relativity has in the form of its diffeomorphism invariance. This symmetry, or rather, the fact that it seems to be an essential feature of general relativity will be the source of the technical and conceptual problems associated with the problem of time. I will argue that this symmetry can be understood as a gauge symmetry, although there is an important difference with gauge theories like electromagnetism: while in these theories one can understand the transformation in a local fashion as leaving invariant the physical content at a time, this version of gauge transformations does not apply for reparametrization invariant theories.

In section \ref{sect_problem_of_time} I will briefly introduce the canonical quantization procedure for gauge theories and I will explain the way in which it gives rise to the problem of time. That is, I will show that the quantum analogs of the constraint equations of any reparametrization invariant system automatically imply that the dynamical equation of the system, i.e., the Schr\"odinger-like equation, becomes trivial. This means that `physical' states are time-independent, which constitutes the problem of time. I will also briefly discuss what I would count as a resolution to this problem.

Then, in section \ref{sect_deparam_and_non-deparam} I introduce the distinction between deparametrizable and non-deparametrizable reparametrization invariant models, which is crucial for my argument. A deparametrizable model is a model in which there is a variable in the configuration or phase space that can be identified with a time variable, or, alternatively, a model in which there is a set of variables in configuration or phase space that can be identified with spacetime coordinates. Conversely, a non-deparametrizable model is a model for which there isn't any such variable. I will give some examples of both kinds of models and I will argue that the problem of time can, in principle, be solved for deparametrizable models, while it is not the case for non-deparametrizable ones.

Finally, in section \ref{sect_failure_of_quantization} I turn to the case of general relativity, I argue that it is a non-deparametrizable theory, and I therefore conclude that the problem of time cannot be solved for the canonical quantization of this theory. That is, contrary to what happens in the case of the deparametrizable models, the apparently timeless states one obtains by canonically quantizing the theory are indeed timeless and problematic.

\section{General relativity and reparametrization invariant models as gauge theories} \label{sect_gauge}

The problem of time arises because of the diffeomorphism of general relativity, and in general for any model with reparametrization invariance\footnote{In this sense, notice that the problem of time is more related to the fact that general relativity is generally covariant than with the fact that it is a gravitational theory. In the quantization of gravitational theories which are not generally covariant there wouldn't appear any problem of time.}. From a technical point of view, these models are expressed in the Hamiltonian form as constrained systems\footnote{I refer the reader to \cite{Rothe2010} for an introduction to the Hamiltonian dynamics of constrained dynamics.}, similarly to gauge theories, and it is for this reason that one follows a similar quantization route. Moreover, from a conceptual point of view reparametrization invariance can be seen as a gauge symmetry, although in this section I will introduce some distinctions, contrary to some claims in the literature.

Let me start by defining reparametrization invariance. A model is reparametrization invariant if its solutions are expressed in terms of some parameters, typically some spacetime coordinates or some parameter parametrizing a trajectory or curve, and, given one solution, one can obtain a physically equivalent solution just by changing the parameters. More technically, the model defines solutions $f_A (\lambda_{\alpha})$, where $f _A$ are the fields or variables in the theory, which vary with respect to a series of parameters $\lambda_{\alpha}$, and which are equivalent under invertible, differentiable transformations from one set of $\lambda_{\alpha}$ to another. These transformations are diffeomorphisms, although I will refer to them as reparametrizations and use the term diffeomorphism just when the parameters $\lambda_{\alpha}$ are coordinates on some spacetime manifold.  

From the point of view of this article, the reparametrization invariance that will be relevant is temporal reparametrization invariance. That is, we will study systems that are described as Lagrangian systems in terms of evolution with respect to a `time' variable or coordinate $\tau$ and which have a symmetry under invertible and differentiable transformations $\tau \rightarrow \tau'$. The case of general relativity is more complicated as there are also spatial diffeomorphisms and combinations of spatial and temporal diffeomorphisms, but for the purposes of this paper it is just temporal reparametrizations that will be playing a role.

Reparametrization invariance is similar in many aspects to the gauge invariance of theories like electromagnetism. Indeed, a reparametrization is a gauge symmetry in the sense that any two solutions of the equations of motion of the theory that are related by a reparametrization are physically equivalent, just as any two solutions of the equations of motion for the 4-potential of electromagnetism which are related by a gauge transformation are physically equivalent. 

In a gauge theory like electromagnetism it is meaningful to speak about the observable quantities at a time $t$ because the gauge transformation doesn't affect the temporal structure of the theory and one can conceive of the gauge transformation as acting at a moment of time. For reparametrization invariant models this is not the case, as the action of the reparametrization is to change the physical state of the world associated with a given time parameter and one cannot interpret this as a transformation leaving unchanged the physical happenings at an instant defined by that time parameter. For instance, two diffeomorphism-related solutions of general relativity could assign the same time parameter $t=t_0$ to two radically different moments of the history of our universe, such as an instant just after the big bang and the instant in which I am typing these words. Clearly, the physical state of the world at such moments was completely different and it doesn't make sense to claim that it is the invariant content under a transformation relating both instants which captures the physical state of the world at time $t_0$.

Arguments along these lines have been given in the literature\footnote{See for instance \cite{PonsSalisbury,Pons2010,Pitts2017,MozotaFrauca2023}. See also the more philosophical discussion in \cite{Maudlin2002} in which this position is defended from the arguments in \cite{Earman2002-EARTMM}. } for reaching the same conclusion I am arriving at here, that reparametrization invariance is a gauge symmetry in the sense that it transforms solutions of the dynamical equations into equivalent solutions, but not in the more local sense, as the physical state at a coordinate time $t$ changes under such a transformation. This difference is relevant at the time of quantizing the theory, as it shows that imposing strict invariance under symmetry transformations would be physically nonsensical in the classical theory, but in the case of the quantum theory this conclusion seems not to be avoidable.   

\section{The problem of time} \label{sect_problem_of_time}

Given the conceptual and formal similarities between reparametrization invariant theories and standard gauge theories it is natural that one attempts to apply the same quantization techniques that work for the latter to the former. However,  here I will show how this leads to a problem of time.

The canonical quantization method for gauge theories was first formulated by Dirac \cite{Dirac1964} and can be summarized as:
\begin{enumerate}
\item{Start with a classical gauge theory defined on a phase space.}
\item{Choose a subalgebra of functions on phase space and quantize them, i.e., build an algebra of operators on a kinematical Hilbert space $\mathcal{H}_{kin}$ such that their commutator algebra is defined by the Poisson algebra of the classical functions.}
\item{Impose the quantum counterparts of the constraints ${\phi}_A$. That is, define the physical Hilbert space $\mathcal{H}_{phys}$ as the space of the states which satisfy $\hat{\phi}_A\vert\psi\rangle=0$.}
\item{Build a Hamiltonian operator which is a quantization of one of the Hamiltonians that generate the constrained dynamics in the classical theory. The dynamics of the theory is contained in the Schr\"odinger equation for that Hamiltonian or in some equivalent form.}
\end{enumerate}
I refer the reader to \cite{MozotaFrauca2023} and the references therein for a detailed discussion of this quantization procedure and for the technical details. For the discussion here we just need to focus on steps 3 and 4 for the case of reparametrization invariant models, as it is in the application of these steps that the problem arises.

In reparametrization invariant models, as in any gauge theory, there are some constraints, but unlike standard gauge theories, the Hamiltonian of the model is itself a constraint\footnote{I refer the reader to \cite{MozotaFrauca2023} for a detailed discussion of the constraints that appear in reparametrization invariant systems and of appropriate Hamiltonian for describing such systems. There are three possible Hamiltonians one can use for describing the system (the canonical, the total, or the extended) but I won't argue here for choosing one of them over the others as the relevant fact is that the three of them are constraints.}. This fact is just a peculiarity in the classical case that doesn't prevent the definition of a consistent classical theory. In particular, it reflects the fact that the Hamiltonian plays not only the role of the generator of time evolutions but also the role of gauge generator in the global sense as described above. In the quantum theory, this peculiarity becomes problematic and gives rise to the problem of time.

Step 3 tells us that the constraints need to be imposed, i.e., that `physical' states satisfy $\hat{\phi}_A\vert\psi\rangle=0$. However, as the Hamiltonian is of the form $H=v^A\phi_A$ this leads to a trivial Schr\"odinger equation:
\begin{equation}
\partial_t \vert \psi \rangle = \hat{H} \vert \psi \rangle =0 \, .
\end{equation}
This means that `physical' states are time-independent, contrary to our expectation from standard quantum theory in which states carry a temporal dependence that allows describing their evolution. In this sense, it seems that we have found a timeless theory in which nothing evolves with time. This is the problem of time for reparametrization invariant theories\footnote{For more extensive and detailed accounts see the classical reviews \cite{Kuchar1992,Isham1993} and the more recent book \cite{Anderson2017}. There are several more or less connected issues that are referred to as `the problem of time', as is well noted in these references. Here I am focusing on what is also known as the frozen formalism problem, i.e., in the fact that the formalism obtained lacks temporal dependence.}.

This is prima facie problematic because it seems that we are missing an important part of what we take to be a quantum theory, its dynamical aspect, and therefore it seems that the quantization has failed. However, in section \ref{sect_deparam_and_non-deparam} I will show how for a class of models, the deparametrizable ones, the problem can be solved in the sense that one is able to interpret states in the physical Hilbert space as dynamical states, i.e., as describing a temporal evolution and recovering the standard quantum-mechanical picture. For the rest of the models, the non-deparametrizable ones, I will argue that it is not the case and that the quantization indeed fails.

Let me clarify that by `solving' the problem of time I don't mean that one recovers a standard quantum theory, as one may very well suspect that the standard structures of quantum theory need to be modified when one attempts to formulate a theory of quantum gravity. A resolution of the problem of time, I argue, consists in showing that a meaningful theory can be built from the `timeless' states obtained in the canonical quantization, which would allow us to claim that the quantization has not failed. This definition of resolution is quite open and it does not presuppose that a meaningful theory is just a theory in the form of a standard quantum theory. Even in this wide sense of resolution, I will argue next that for non-deparametrizable models we lack a successful resolution, and hence that the canonical quantization of these models fails.

\section{Deparametrizable and non-deparametrizable systems} \label{sect_deparam_and_non-deparam}

As I just mentioned, the main feature of a reparametrization invariant theory that determines whether the problem of time can be solved or not is whether it is deparametrizable or not. In this section I will give a definition and examples of both deparametrizable and non-deparametrizable systems and I will argue why the problem of time can be solved for the former but not for the latter.

Let me start by giving the definitions of both types of reparametrization invariant systems:

\textbf{Deparametrizable model:} A model is deparametrizable if one can identify a time variable or a set of spacetime coordinates among its configuration or phase space variables.

\textbf{Non-deparametrizable model:} A model is non-deparametrizable if it is not deparametrizable.

A deparametrizable model is one that can be deparametrized, i.e., it is possible to express evolution not with respect to the arbitrary parameter(s) of the reparametrization invariant model, but with respect to the time variable, or spacetime coordinates, that can be found among the variables of the model. For instance, consider the model given by the following action:
\begin{equation} \label{deparametrizable_Newtonian_body}
S[x,t]=\int d\tau \left(\frac{1}{2}m\frac{\dot{x}^2}{\dot{t}}-\dot{t}V(x)\right) \, .
\end{equation}
Solutions of the equations of motion for this action are of the form $x(\tau),t(\tau)$ and describe the trajectory of a body in Newtonian spacetime under the effect of the potential $V$. The trajectory is independent of the way it is parametrized and it can be deparametrized and expressed with respect to the true time variable $t$. That is, one can express the trajectory just as $x(t)$. 

A necessary condition for deparametrization is that the variable to be identified as a time variable needs to be monotonic with respect to the arbitrary parameter $\tau$. However, it is not a sufficient condition, as even in models in which there are physical variables that are monotonic they can be argued to be different from a time variable. For instance, the position of a particle is monotonic in Newtonian physics if it is a free particle, but it is different from time as we can conceive a situation in which it is affected by some potential and ceases to be monotonic. In this sense, my definition of deparametrizable is not purely formal regarding monotonic functions, but has a component of how we interpret some variables as time variables or spacetime coordinates.

\begin{figure}[h]%
\centering
\includegraphics[width=0.9\textwidth]{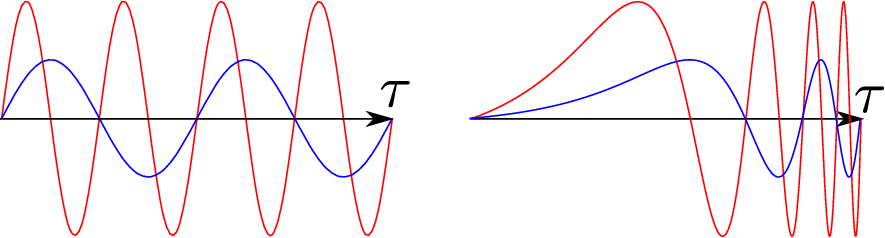}
\caption{\label{oscillators} Two equivalent solutions of the equations of motion of the double harmonic oscillator model. They represent the same sequence of oscillations but they differ from each other at the particular values of $\tau$ they assign to each moment of time. In this way the parametrization on the left-hand side represents the Newtonian parametrization in which the oscillations are regular and the one on the right-hand side represents a parametrization in which the oscillations become faster as $\tau$ passes.}
\end{figure}

For the case of non-deparametrizable models there is no variable that plays the role of time (or spacetime coordinates) in the configuration or phase space of theory and one cannot perform a deparametrization. The most popular example of such a model is the Jacobi action for any Newtonian system\footnote{The relevance of this kind of model for the discussion of the problem of time was first noticed by Julian Barbour \cite{Barbour1994}. The discussion in \cite{Gryb2010} brought the topic again to the fore.}. As an example, let me introduce the Jacobi action for a system of two harmonic oscillators:
\begin{equation} \label{harmonic_osc_reparametrizable}
S[x,y]=2\int d\tau \sqrt{\frac{m}{2}\left(\dot{x}^2+\dot{y}^2\right)\left(E-\frac{1}{2}(k_xx^2+k_yy^2)\right)} \, .
\end{equation}
Solutions of the equations of motion of this system are pairs $x(\tau),y(\tau)$ which describe the trajectories of the two oscillators with respect to an arbitrary temporal parametrization $\tau$. In figure \ref{oscillators} I represent two equivalent such parametrizations that represent the same physics, i.e., the same sequence of configurations. Note that neither of the configuration space variables is a time variable, as both have a different physical interpretation and furthermore none of them is monotonic.

To recover a preferred notion of time one can `gauge transform' to Newtonian time $t$ by means of the relation $dt=\sqrt{\frac{m(\dot{x}^2+\dot{y}^2)}{2E-k_xx^2-k_yy^2}}d\tau$ that relates Newtonian time with the parameter $\tau$ describing the same trajectory using a different parametrization. In this sense, it is important to emphasize that the physical content of the model is encoded in the trajectories $x(\tau),y(\tau)$ or $x(t),y(t)$, but not on some `deparametrized' $x(y)$ or $y(x)$. To insist, while in the deparametrizable case we could identify one configuration variable as time, this is not the case in the non-deparametrizable case.

This fact is determinant for the different ways the problem of time affects both kinds of models. For the case of deparametrizable models, states in the physical Hilbert space can be interpreted, in one way or another, as describing dynamical states with respect to a time variable and therefore the problem is solved. I refer the reader to \cite{MozotaFrauca2023} for a detailed discussion of the different strategies proposed for solving the problem in this case and for an argument showing that they all rely on a deparametrization. Here I will give just a simple account of the example above that can be generalized to more complicated systems and to different strategies of resolution.

Applying the quantization steps in section \ref{sect_problem_of_time} to the deparametrizable system described by action \ref{deparametrizable_Newtonian_body} we are lead to a problem of time, as we find states of the form $\psi(x,t)$ which do not evolve with respect to $\tau$. However, this is not problematic as we can nevertheless interpret $\psi(x,t)$ not as a state at `time' $\tau$ but as a wavefunction for the true configuration variable $x$ which evolves with respect to time $t$. Moreover, the quantum constraint equation for this system is just the Schr\"odinger equation and one can interpret states in the physical Hilbert space as solutions to the dynamical equation of the quantum system. In this sense, the problem of time can be solved for this system: the apparently timeless states were hiding just standard quantum mechanics.

However, the case for non-deparametrizable models is not so positive. Consider the case of the double oscillator above. `Physical' states are states of the form $\psi(x,y)$ and, contrary to what happened in the deparametrizable case, they are clearly not the dynamical states one would expect from the non-reparametrization invariant form of the model, which would be states of the form $\psi(x,y,t)$ where $t$ is the Newtonian time variable. One could try to look for an interpretation that would nevertheless interpret $\psi(x,y)$ as states containing dynamical information and from which a meaningful theory could be devised. For instance, one could try to mimic the interpretation for the deparametrizable system and interpret $\psi(x,y)$ as a wavefunction on the configuration space for $x$ that evolves with respect to the `time' $y$. However, this is problematic for several reasons. First, it breaks the symmetry between $x$ and $y$ as both of them play the same role in the classical theory and receive the same interpretation. Second, it seems wrong to identify $y$ as a time variable as it clearly wasn't a time variable in the classical theory and as it wasn't even monotonic. In the classical theory $y$ was oscillating in time and a value $y_0$ couldn't be used to identify a particular instant of the trajectory, as there were many, infinitely many indeed, instants in which the oscillator took the position $y_0$. In the quantum theory there would be just one wavefunction $\psi(x)$ for `time' $y_0$ which wouldn't be reflecting the oscillating behavior of the oscillator. Furthermore, if we read $y$ as a time variable we would take $y\rightarrow -\infty$ as the distant past and $y\rightarrow \infty$ as the distant future in the quantum theory, while in the classical theory the values of $y$ were bounded and oscillating. Moreover, the relationship between the quantum theory we have obtained and the classical theory is unclear to say the least, as it seems that we cannot recover the behavior of the classical theory starting from the quantum one.

Similar objections can be raised to the different strategies for finding a resolution that have been proposed in the literature. That is, when applied to the model of the two harmonic oscillators, the strategies that work for deparametrizable models can be argued to fail. This is mainly because these strategies need in a more or less explicit way a deparametrization, and when it is not available one runs into trouble. I refer the reader to \cite{MozotaFrauca2023} for a detailed discussion of the different strategies and the way they seem to fail in the case of the double harmonic oscillator. 

Let me also mention that one could try to devise a Bohmian strategy in which even if the state is time-independent this wouldn't be problematic if the dynamics of the Bohmian particles or fundamental entities is non-trivial. However, the guidance equation would need to be changed and I am unaware of any generic proposal for solving this for any reparametrization invariant system. There are some proposals for the case of the quantization of general relativity, but they aren't popular in the quantum gravity literature\footnote{See \cite{Pinto-Neto2018} and references therein.}.

For these reasons, I have argued that the canonical quantization of non-deparametrizable models fails, or, at least, that the proposed resolutions for the problem of time for these models have serious shortcomings that would need to be addressed in order to consider that they show how the timeless states obtained by canonically quantizing these models could encode a meaningful theory.

\section{Does the quantization fail for general relativity?} \label{sect_failure_of_quantization}

Finally, we can turn to the case of general relativity. As I have advanced before, I will argue that general relativity is not deparametrizable and therefore that the problem of time for its canonical quantization cannot be solved as it was solved in the case of deparametrizable models.

For the canonical quantization procedure, general relativity is expressed as a Lagrangian theory in which the geometry of space and matter evolve with respect to an arbitrary time parameter, that is, a foliation is introduced that splits the relativistic spacetime into space and time\footnote{Indeed, not every relativistic spacetime can be decomposed in this way, and they are left out for the quantization of the theory.}. The degrees of freedom of the theory are a three-dimensional metric $g_{ab}$ that describes the geometry of space, the lapse function $N$ and shift vector $N^a$ that encode the time-like components of the metric, and possibly some degrees of freedom corresponding to matter. Under a diffeomorphism, these degrees of freedom are affected, but we know that the diffeomorphism-related description is just an equivalent description of the same spacetime. For instance, one can have two diffeomorphism-related descriptions of Minkowski spacetime, one in which time slices describe flat spaces and another one in which they don't.

A priori, all the variables in this description have a physical interpretation that is different from being a time variable or a spacetime coordinate. In this sense, if general relativity were deparametrizable, it would not be in a straightforward way. Some authors proposed some ways in which spacetime geometry could be encoding time in some way which could be used for claiming that general relativity is deparametrizable after all\footnote{See for instance \cite{Baierlein1962,Kiefer2012}.}. However, the consensus about this issue is that general relativity isn't deparametrizable, as it was argued for instance in \cite{Kuchar1992}. One of the reasons for this is that it seems quite clear that there is no phase space function that satisfies the necessary condition of being monotonic. In \cite{MozotaFrauca2023} I further counter-argued against some arguments in the literature used for supporting the claim that general relativity is deparametrizable. My analysis of these arguments shows that they could be applied to the case of the double harmonic oscillator and that they would lead to the wrong conclusion that this system is deparametrizable. As these arguments aren't valid when applied to the case of the double harmonic oscillator model, I argued that they aren't valid when applied to general relativity.

There is a further argument linking the structure of general relativity to the structure of our example above. The definition of proper time in general relativity is analogous to the definition of Newtonian time in the model defined in the Jacobi action. In particular, proper time is defined by:
\begin{equation}
ds^2=-g_{\mu\nu}dx^{\mu}dx^{\nu}=N^2dt^2-g_{ab}(dx^a+N^adt)(dx^b+N^bdt) \, ,
\end{equation}
which is just analogous to $dt=\sqrt{\frac{m(\dot{x}^2+\dot{y}^2)}{2E-k_xx^2-k_yy^2}}d\tau$ for the double oscillator. For these reasons, there are good grounds for comparing general relativity with the example of the double harmonic oscillator and arguing that if the quantization fails for one of the two, then the quantization must fail for the other.

This conclusion puts some pressure on the quantum gravity community, at least on those working on approaches based on canonical quantizations of general relativity. In particular, this affects approaches like quantum geometrodynamics, loop quantum gravity, loop quantum cosmology, and some other approaches to quantum cosmology. There are a couple of ways the community could react to defend these approaches: they could either deny that general relativity is a non-deparametrizable model, or argue that it is of a different kind from the example in this paper, so that the comparison doesn't affect their theories; or they could argue that there is a resolution of the problem of time that is satisfactory for both general relativity and other non-deparametrizable models. In any case, the defender of these approaches would need to clarify the way in which they are believed to constitute satisfactory quantum theories.

\section*{Acknowledgments}

This research is part of the Proteus project that has received funding from the European Research Council (ERC) under the  Horizon 2020 research and innovation programme (Grant agreement No. 758145) and of the project CHRONOS (PID2019-108762GB-I00) of the Spanish Ministry of Science and Innovation.

\end{document}